\documentclass[prd,twocolumn,floatfix,preprintnumbers,letterpaper]{revtex4}
\usepackage{graphics}
\usepackage{epsfig}
\usepackage{amsmath}

\begin{document}

\title{Phantom Dark Energy, Cosmic Doomsday, and the Coincidence Problem}
\author{Robert J. Scherrer}
\affiliation{Department of Physics and Astronomy, Vanderbilt University,
Nashville, TN  ~~37235}

\begin{abstract}
Phantom dark energy models, with $w < -1$, are characterized by a future singularity and
therefore a finite
lifetime for the universe.  Because the future singularity is triggered by
the onset of dark-energy domination, the universe spends a significant fraction of its total
lifetime in a state for which the dark energy and matter densities are roughly comparable.
We calculate, as a function of $w$,
the fraction of the total lifetime of the universe for which the dark energy and
matter
densities differ by less than the ratio $r_0$ in either direction.  For $r_0$ = 10, this
fraction varies from $1/3$ to $1/8$ as $w$ varies from $-1.5$ to $-1.1$; the fraction is smaller
for smaller values of $r_0$.  This result indicates that the coincidence problem is significantly
ameliorated in phantom-dominated cosmologies with a future singularity.

\end{abstract}

\maketitle

\section{Introduction}

The universe appears to consist of approximately 30\% nonrelativistic matter, including
both baryons and dark matter, and
70\% dark
energy (see Ref.
\cite{Sahni} for a recent review, and references therein).  The matter has a density $\rho_M$
that scales with the scale factor $R$ as $\rho_M \propto R^{-3}$.  The evolution of the
dark energy density depends on its equation of state, which
is usually parametrized in the form
\begin{equation}
p_{DE} = w \rho_{DE},
\end{equation}
where $p_{DE}$ and $\rho_{DE}$ are the pressure and density of the dark energy.
Then the density of the dark energy scales as
\begin{equation}
\label{rhoDE}
\rho_{DE} \propto R^{-3(1+w)}.
\end{equation}
The simplest model for the dark energy is a cosmological constant, for which $w = -1$ and
$\rho_{DE} = constant$.
More complex models have been proposed, in which the dark energy arises from a scalar field; these
are called quintessence models \cite{ratra,turner,caldwelletal,liddle,zlatev}.  These models generally give rise to a time-varying
$w$ and more complex behavior for $\rho_{DE}$.

However, all of these models have a similar problem.  Since the matter density decreases more
rapidly than the dark energy density, we expect $\rho_M \gg \rho_{DE}$
at times much earlier than the present, while
in the far future, $\rho_{DE} \gg \rho_M$.  Thus, we live in a special epoch, when the matter
and dark energy densities are roughly comparable.  This has been dubbed the ``coincidence problem."

It is possible, of course, that this coincidence is, in fact, just a coincidence,
and no deeper explanation is possible.
Nonetheless, numerous attempts have been made to provide an explanation for this coincidence.
For example, in a certain class of quintessence models with non-standard kinetic terms (``$k$-essence"),
the dark energy density tracks the
radiation energy density during the radiation-dominated era, but then evolves toward a
constant-density dark energy component during the matter-dominated era \cite{Arm2,Arm3},
so the coincidence problem is resolved by linking the
onset of dark energy domination to the epoch of equal matter and radiation.  (Although
see Malquarti et al. \cite{Malquarti1}, who argued
that the basin of attraction for models exhibiting the desired behavior is quite
small).  Another possible solution lies with models in which the periods of matter
domination and dark energy domination alternate, either because the dark energy arises from a single
scalar field with oscillatory behavior \cite{Dodelson,quintom}, or because the periods of dark energy domination
arise from different quintessence fields, with a variety of energy scales \cite{Griest}.
Another possibility is to couple the matter and quintessence fields, so that energy can be transferred between
them.  With a suitable coupling, a constant ratio of dark energy density to matter density
can be arranged \cite{Chimento1,Chimento2,Huey}.  Finally, anthropic arguments have been put forward to resolve the
coincidence problem (see, e.g., Ref. \cite{Garriga}), but such arguments are, of course,
controversial.

The coincidence problem is significantly ameloriated in phantom dark energy cosmologies, first
proposed by Caldwell \cite{Caldwell}.  Such models, with $w < -1$, can lead to a future singularity,
or ``cosmic doomsday" \cite{rip,rip2}. (Note that Starobinsky \cite{Star} previously considered
similar models and calculated the time to the future singularity).  The existence
of this future singularity is often considered a negative feature of phantom dark energy models, so a great deal of
effort has gone into constructing models with $w<-1$ that avoid a future singularity
\cite{Diaz,Sahni2,Nojiri1,Nojiri2,Nojiri3,Woodard,Yu}.  Here we examine a positive feature of phantom dark energy
models with a future singularity, previously mentioned in passing by McInnes \cite{mcinnes2}:
since such models produce a finite lifetime for the universe, it is possible to
calculate the fraction of time that the universe spends in a ``coincidental state", with $\rho_{M} \approx
\rho_{DE}$.  Because the final singularity is triggered by the onset of dark energy domination,
this fraction is generically nonnegligible.  Here we make this argument quantitative.

In the next section, we examine the time evolution of phantom cosmologies, and calculate the fraction of the total
lifetime of the universe for which the matter and dark energy densities are roughly similar.  Our results are
discussed in Sec. III.

\section{Phantom Cosmologies and the Coincidental State}

Our goal is to determine, for a universe
containing phantom dark energy, the fraction of the total lifetime of the universe for
which $\rho_{DE} \approx \rho_{M}$.  Obviously, the question of what present-day observed
value of $\rho_{DE}/\rho_{M}$ would constitute a ``coincidence" is not well-defined.  The currently
observed value is roughly 2, but a ratio within an order of magnitude, i.e.,
$1/10 < \rho_{DE}/\rho_{M}< 10$ would certainly be deemed ``coincidental."  Hence,
we leave the definition of a coincidence as a free parameter in the calculation.  Specifically,
we define the parameter $r$ to be the ratio of the dark energy and matter densities:
\begin{equation}
\label{rdef}
r \equiv \frac{\rho_{DE}}{\rho_{M}} = \frac{\rho_{DE0}}{\rho_{M0}}(R/R_0)^{-3w},
\end{equation}
where $\rho_{M0}$ and $\rho_{DE0}$ are the matter and dark energy densities evaluated at
an arbitrary fiducial scale factor $R_0$. 
Then for any fixed ratio $r_0$, we can calculate the fraction of time for which
$1/r_0 < r < r_0$, i.e., the ratios of the dark energy and matter densities are within a factor
of $r_0$ of each other in either direction.

Phantom cosmologies are characterized by a dark energy component with $w < -1$ \cite{Caldwell}.
Recent supernova limits on $w$ give $w \gtrsim -1.5$ \cite{Knop,Riess}, so we will
take this as a conservative lower bound on $w$.
Since the
dark energy density scales as in equation (\ref{rhoDE}), the density of dark energy increases with the
expansion of the universe.  To see why this leads to a future singularity,
consider a model containing both matter and phantom dark energy.  We will assume,
for simplicity, that $w$ for the phantom dark energy is constant, but our results generalize in an obvious way to
non-constant $w$.  The Friedmann equation for a mixture of matter and phantom dark energy is
\begin{equation}
\label{Fried}
\left(\frac{\dot R}{R}\right)^2 = \frac{8}{3}\pi G \left( \rho_{M0}(R/R_0)^{-3}
+ \rho_{DE0}(R/R_0)^{-3(1+w)} \right).
\end{equation}
When $\rho_{M} \ll \rho_{DE}$, equation (\ref{Fried})
can be integrated to give
\cite{Caldwell}
\begin{equation}
R(t) = R(t_m)[-w + (1+w)(t/t_m)]^{2/3(1+w)},
\end{equation}
where $t_m$ is the time at which the matter and phantom dark energy densities are equal.
Thus, $R$ and $\rho_{DE}$ both go to infinity at a finite time, $t_U$,
given by
\begin{equation}
\label{tfin}
t_U = \left(\frac{w}{1+w}\right)t_m.
\end{equation}

With regard to the coincidence problem, there are two key points.  First, it is the onset
of dark energy domination which triggers this ``cosmic doomsday", so $t_m$ is naturally a non-negligible
fraction of $t_U$, as shown in equation (\ref{tfin}).  Second,
at the other end of the time scale, because of the power
law nature of the expansion, the time during which the universe expands
from $r = 1/r_0$ to $r = r_0$ will be larger than
the time in expanding from $r = 0$ to $r = 1/r_0$, as long as $r_0$ is sufficiently large.
These two results indicate that
the phantom dark energy universe naturally spends a significant fraction of its lifetime in the state for which
$\rho_{M}$ and $\rho_{DE}$ are roughly comparable.

The total lifetime $t_U$ of the phantom dark energy universe can be calculated more accurately
by integrating equation (\ref{Fried}):
\begin{eqnarray}
\label{tU}
t_U &=& \int_0^\infty   R^{-1} \biggl[\frac{8}{3}\pi G \biggl(\rho_{M0}(R/R_0)^{-3} \nonumber\\
&+& \rho_{DE0}(R/R_0)^{-3(1+w)}\biggr) \biggr] ^{-1/2} dR.
\end{eqnarray}
The total time that the universe spends in expanding from
a given initial scale factor, $R_1$, to a given final scale factor $R_2$, is:
\begin{eqnarray}
\label{tR}
t_{12} &=& \int_{R_1}^{R_2}  R^{-1}\biggl[\frac{8}{3}\pi G\biggl( \rho_{M0}(R/R_0)^{-3}\nonumber\\
&+& \rho_{DE0}(R/R_0)^{-3(1+w)} \biggr)\biggr] ^{-1/2} dR.
\end{eqnarray}
Then the fraction $f$ of the total lifetime of the universe that it spends in expanding
from $R_1$ to $R_2$ is just $t_{12}/t_U$.
Since we are interested in the time over which $\rho_{M} \approx \rho_{DE}$, it
makes sense to express
$R_1$ and $R_2$ in terms of the density ratio $r$ defined in equation (\ref{rdef}).
Rewriting equations (\ref{tU}) and (\ref{tR}) in terms of $r$, rather than $R$,
we get a simple expression for $f =  t_{12}/t_U$:
\begin{equation}
f = \frac{\int_{r_1}^{r_2} r^{-(2w+1)/2w}/\sqrt{1+r} ~dr}{\int_{0}^{\infty} r^{-(2w+1)/2w}/\sqrt{1+r} ~dr}.
\end{equation}
The integral in the denominator converges slowly for $w$ near $-1$, but it can be expressed in terms
of gamma functions, giving:
\begin{equation}
\label{fr}
f = \frac{\Gamma(1/2)}{\Gamma(-\frac{1}{2w})\Gamma(\frac{1}{2} + \frac{1}{2w})}\int_{r_1}^{r_2} \frac{r^{-(2w+1)/2w}}
{\sqrt{1+r}} dr.
\end{equation}
Equation (\ref{fr}) provides the main result of this paper; it tells us what fraction $f$
of the total (finite) lifetime
of a phantom universe is spent in a ``coincidental" state, with $\rho_{DE}/\rho_{M}$ between $r_1$
and $r_2$.  As expected, $f$ is a function only of $r_1$, $r_2$, and $w$; it is independent of the current
matter and dark energy densities.

As noted above,
we choose to calculate $f$ as a function of the parameter $r_0$,
taking the limits $r_1 = 1/r_0$ and $r_2 = r_0$.  This corresponds to determining the fraction of time
for which the matter and dark energy densities are within a factor of $r_0$ of each other in either direction.
In Fig. 1, we show $f$ as a function of $r_0$ for some representative values of $w$. 
\begin{figure}[htb]
\centerline{\epsfxsize=3truein\epsffile{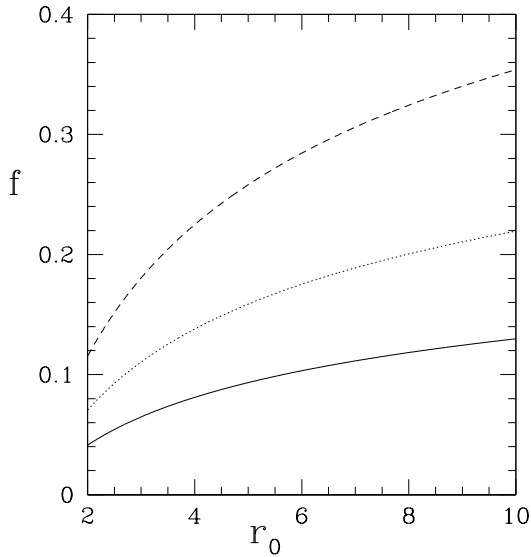}}
\caption{The fraction of time $f$ that a phantom cosmology spends in a ``coincidental" state, defined
as $1/r_0 < \rho_{DE}/\rho_{M} < r_0$, for $w = -1.1$ (solid curve), $w=-1.2$ (dotted curve),
and $w = -1.5$ (dashed curve).}
\end{figure}

As we have suggested, the total fraction of time that phantom cosmologies spend in a ``coincidental" state
can be quite large, and it increases as $w$ becomes more negative.  If we take a ``coincidence" to be a state
for which $\rho_{M}$ and $\rho_{DE}$ differ by less than a factor of 10 (in either direction), then
a universe with $w=-1.5$ spends more than a third of its lifetime in such a state.  This fraction decreases
to roughly 1/5 for $w=-1.2$ and  approximately 1/8 for  $w=-1.1$; these are still substantial fractions of the
total lifetime of the universe.  If we apply a more stringent standard for what constitutes a ``coincidence",
then these fractions obviously shrink.  For instance, the fraction of time for which
$\rho_{M}$ and $\rho_{DE}$ differ by less than a factor of 2 varies from 4\% for $w=-1.1$ up to 12\% for $w=-1.5$.  Even so, these are not
negligible fractions of the total lifetime of the universe.

\section{Discussion}

Our results indicate that a universe with phantom dark energy spends a significant fraction of its total
lifetime in a state for which $\rho_{M}$ and $\rho_{DE}$ differ by less than an order of magnitude.  Thus,
if we simply found ourselves living at a random moment in such a universe, it would not be surprising
to find $\rho_{M}$ and $\rho_{DE}$ in such a ``coincidental" state.
Clearly, these results do not explain {\it why} we happen to live in the epoch for which $\rho_{M} \approx \rho_{DE}$;
instead, they make this possibility less implausible.
These arguments generalize, in an obvious way, to phantom models in which $w$ varies with time.

Note that this is
{\it not} an anthropic argument; it does not rely on the universe being in a state which can support
life.  For instance, this argument could also be applied to a model of the universe
in which the ratio of the dark energy, matter, and radiation densities was such that
the universe passed directly from the radiation-dominated era to the dark-energy dominated era, without
an intervening matter-dominated era.  From an anthropic point of view, such a universe is ruled out
by the lack of structure formation.  In contrast, the argument presented here would simply indicate that
the universe in this case
would spend a significant fraction of its lifetime with the radiation and dark energy
densities having the same order of magnitude.  Caldwell et al. \cite{rip} noted that
the phantom dark energy cosmology forces us to live in the brief epoch in which bound structures can form, but
this again is a different argument from the one we are making here.

These results clearly depend on the fact that a phantom-dominated universe has a finite lifetime.  The only way
to calculate an equivalent fraction of time in an infinitely long-lived universe would be if
the epochs of dark energy domination and matter domination alternated in a cyclical fashion, as in the models
of Refs. \cite{Dodelson,quintom,Griest}.  In such models one could then determine the probability of living in a ``coincidental
epoch" of the sort in which we presently find ourselves.

\acknowledgments

R.J.S. was supported in part by the Department of Energy (DE-FG05-85ER40226).

\newcommand\AJ[3]{~Astron. J.{\bf ~#1}, #2~(#3)}
\newcommand\APJ[3]{~Astrophys. J.{\bf ~#1}, #2~ (#3)}
\newcommand\apjl[3]{~Astrophys. J. Lett. {\bf ~#1}, L#2~(#3)}
\newcommand\ass[3]{~Astrophys. Space Sci.{\bf ~#1}, #2~(#3)}
\newcommand\cqg[3]{~Class. Quant. Grav.{\bf ~#1}, #2~(#3)}
\newcommand\mnras[3]{~Mon. Not. R. Astron. Soc.{\bf ~#1}, #2~(#3)}
\newcommand\mpla[3]{~Mod. Phys. Lett. A{\bf ~#1}, #2~(#3)}
\newcommand\npb[3]{~Nucl. Phys. B{\bf ~#1}, #2~(#3)}
\newcommand\plb[3]{~Phys. Lett. B{\bf ~#1}, #2~(#3)}
\newcommand\pr[3]{~Phys. Rev.{\bf ~#1}, #2~(#3)}
\newcommand\PRL[3]{~Phys. Rev. Lett.{\bf ~#1}, #2~(#3)}
\newcommand\PRD[3]{~Phys. Rev. D{\bf ~#1}, #2~(#3)}
\newcommand\prog[3]{~Prog. Theor. Phys.{\bf ~#1}, #2~(#3)}
\newcommand\RMP[3]{~Rev. Mod. Phys.{\bf ~#1}, #2~(#3)}

\end{document}